\documentstyle[aps]{revtex}
\newcommand{\be}{\begin{equation}}
\newcommand{\ee}{\end{equation}}
\newcommand{\ben}{\begin{eqnarray}}
\newcommand{\een}{\end{eqnarray}}
\begin{document}
 
\title{Exploring the Vicinity of the Bogomol'nyi-Prasad-Sommerfield Bound} 
\author{C A G Almeida, D Bazeia and L Losano} 
\address{Departamento de F\'\i sica, Universidade Federal da Para\'\i ba,\\ 
Caixa Postal 5008, 58051-970 Jo\~ao Pessoa, Para\'\i ba, Brazil} 
\date{\today} 

\maketitle

\begin{abstract}
We investigate systems of real scalar fields
in bidimensional spacetime, dealing with potentials that are small
modifications of potentials that admit supersymmetric extensions. The
modifications are controlled by a real parameter, which allows implementing
a perturbation procedure when such parameter is small. The procedure allows
obtaining the energy and topological charge in closed forms, up to first
order in the parameter. We illustrate the procedure with some examples. In
particular, we show how to remove the degeneracy in energy for the one-field
and the two-field solutions that appear in a model of two real scalar
fields.
\end{abstract}

%%%%%%%%%%%%%%%%%%%%%%%%%%%%%%%%%%%%%%%%%%%%%%%%%%%%%%%%%%%%%%%%%%%%%%%
\section{Introduction}

Domain walls are defect structures that appear in systems engendering
spontaneous breaking of discrete symmetry. They are of interest for instance 
in Condensed Matter, as interfaces in magnetic materials \cite{esc81}, as
seeds for pattern formation \cite{wal97}, and as interfaces in ferroelectric
crystals \cite{jsh93,sal93,sle98}, and in Cosmology, as seeds for the
formation of structures \cite{ktu90,vil94} in the early universe. Domain
walls spring through the immersion of kink-like solutions of $(1,1)$
space-time dimensions to higher spatial dimensions. Standardly, kinks or
domain walls appear in scalar field models, but they may also be present
in extended systems, that include fermionic fields, with or without
supersymmetry.

Recently, the study of supersymmetric models has brought new issues, as for
instance in the investigations of a Wess-Zumino model \cite{gto99,cht00}
engendering the $Z_3$ symmetry. This model allows the identification of
a Bogomol'nyi equation \cite{bog76} for a triple junction that breaks $1/4$
supersymmetry of the model. We learn from these works that supersymmetry
helps easing calculations concerning the presence of the triple
junction \cite{oin99,ato91}. However, supersymmetry seems to play no central
role when the issue is the tiling of the plane and not the triple junction
ifself \cite{saf99,bbl00,bbd00}. Furthermore, in Ref.~{\cite{bbd00r}} it was
shown how to entrap a planar regular hexagonal network of defects inside a
domain wall. This model involves three real scalar fields, and engenders the
$Z_2\times Z_3$ symmetry, but it seems to have no supersymmetric extension.

These facts motivate the study of bosonic models that do not support
supersymmetric extensions. However, to keep track of supersymmetry we
investigate models that are close to the real bosonic portions of
supersymmetric systems. These models are defined by potentials that contain
two parts: the first part defines the real bosonic portion of a
supersymmetric system; the second part defines the extended model.
The first part alone constitutes the basic model, which supports static field
configurations that minimize the energy, attained by static field
configurations that solve first order differential equations. In particular,
we examine models that support topological solutions that belong to the
same topological sector, and are degenerate, having the very same energy. An
example of this was investigated in \cite{bnr97},
in a model of two coupled scalar fields. This model has been recently extended
to the case of several scalar fields in \cite{ilg00}. We explore the
possibility of extending this system, in order to remove the degeneracy
between the one-field and the two-field solutions. This investigation is of
intrinsic interest, and may also help examining applications to Cosmology and
to Condensed Matter. In Cosmology we recall the usual route \cite{ktu90,vil94},
and also the new possibility \cite{bbd00r}. In Condensed Matter, we
envisage other issues, concerning for instance the presence of Ising and
Bloch wall interfaces in magnetic materials described by the anysotropic XY
model \cite{wal97,bbd00r}, and the structural phase transition in
ferroelectric crystals \cite{sle98,cm}.

The real bosonic sector of supersymmetric systems described by $n$ chiral
superfields $\Phi^1_{c},\Phi^2_{c},...,\Phi^n_{c}$ contains $n$ real
scalar fields $\phi_1,\phi_2,...,\phi_n$. The potential $V$ is written in
terms of the superpotential $W$, in a way such that for
$V=V(\phi_1,\phi_2,...,\phi_n)$ and $W=W(\phi_1,\phi_2,...,\phi_n)$ we get
\be
\label{pot}
V=\frac12W^2_{\phi_1}+\frac12W^2_{\phi_2}+\cdots+\frac12W^2_{\phi_n}
\ee
where $W_{\phi_i}=\partial W/\partial\phi_i, i=1,2,...,n$. These systems
have been investigated in several different contexts in \cite{bbd00,bnr97,cm},
and in references therein. In the present work we investigate systems
where the potential includes an extra piece, that
modifies the above potential according to
\be
\label{pote}
V_{\epsilon}=V+\frac12\;\epsilon\;F
\ee
where $\epsilon$ is a parameter, real, and $F=F(\phi_1,\phi_2,...,\phi_n)
$ is in principle an arbitrary function of the fields. In the present work
we show that if $\epsilon$ is small, we can develop a perturbation procedure
that gives closed results up to first order in $\epsilon$. The perturbation
procedure is based on former investigations \cite{baz91,jkh91}, and works
nicely for potentials of the form (\ref{pote}) when the function $F$ obeys
some restrictions. We start our investigations using natural units, but we
work with dimensionless fields and coordinates, and sometimes we refer to
the model with $\epsilon=0$ as the primary system, and to the complete model,
with $\epsilon\neq0$, as the extended system. 

%%%%%%%%%%%%%%%%%%%%%%%%%%%%%%%%%%%%%%%%%%%%%%%%%%%%%%%%%%%%%%%%%%%%%%%%%%%
\section{General Considerations}
\label{sec:general}

In order to better understand the procedure, let us first consider models
that are bosonic portions of supersymmetric theories. In this case we have
\be
{\cal L}=\frac12\sum_{i=1}^n\partial_{\alpha}\phi_i\partial^{\alpha}\phi_i
-V(\phi_1,...,\phi_n)
\ee
where $V$ is given by Eq.~(\ref{pot}). We work in $(1,1)$ space-time
dimensions. The equation of motion for static fields are
\begin{eqnarray*}
& &\frac{d^2\phi_1}{dx^2}=
W_{\phi_1}W_{\phi_1\phi_1}+...+W_{\phi_n}W_{\phi_n\phi_1}
\\
& &\frac{d^2\phi_2}{dx^2}=
W_{\phi_1}W_{\phi_1\phi_2}+...+W_{\phi_n}W_{\phi_n\phi_2}
\\
& &\quad\vdots
\\
& &\frac{d^2\phi_n}{dx^2}=
W_{\phi_1}W_{\phi_1\phi_n}+...+W_{\phi_n}W_{\phi_n\phi_n}
\end{eqnarray*}
These second order equations are solved by field configurations that solve
the first order differential equations
$$
\frac{d\phi_1}{dx}=W_{\phi_1},\;
\frac{d\phi_2}{dx}=W_{\phi_2},\; ...\;\frac{d\phi_n}{dx}=W_{\phi_n}
$$
These are the Bogomol'nyi equations. The energy of the static solutions
can be written as
\be
E=\frac12\int^{\infty}_{-\infty}dx\sum_{i=1}^n
\Biggl[\left(\frac{d\phi_i}{dx}\right)^2+
W^2_{\phi_i}\Biggr]
\ee
In general, the system may have several distinct sectors, which may be
identified by the two vacuum states the static solution connect.
Thus, if we use the set of numbers $\{v_a,v_b,v_c,...\}$ to mark the vacuum
states of the model, we can write
\be
E^{ab}=E^{ab}_B+\frac12\int^{\infty}_{-\infty}dx\sum_{i=1}^n
\left(\frac{d\phi_i}{dx}-W_{\phi_i}\right)^2
\ee
Here $E^{ab}_B=|W_{ab}|$, and $W_{ab}=W(v_a)-W(v_b)$.
A pair of vacua defines a topological sector,
and in the sector $(ab)$ the energy is minimized
to the bound $E^{ab}_B$ for field configurations that obeys the above first
order equations. Solutions of the Bogomol'nyi equations are named
Bogomol'nyi-Prasad-Sommerfield solutions \cite{bog76,pso75}, and we refer
to this bound as the BPS bound. It is possible that $W(v_a)=W(v_b)$, giving
a vanishing $W_{ab}$. In this case the topological sector cannot support BPS
states, and we refer to this as a non-BPS sector -- see Ref.~{\cite{bbd00}}.
The topological features of these solutions can be accounted for by
introducing for instance the topological current \cite{bbl00}
\be
j^{\alpha}=\frac12\;\varepsilon^{\alpha\beta}\partial_{\beta}\Phi
\ee
where $\Phi$ is a column vector, such that
$\Phi^t=(\phi_1\;\phi_2\cdots\phi_n)$.

We illustrate the general situation with some examples. We work with natural
units. However, we avoid unimportant considerations by
considering the Lagrangian density ${\cal L}$, written in terms of
dimensionless fields and coordinates. The usual form of the Lagrangian
density is ${\cal L'}=\gamma\,{\cal L}$, where $\gamma$ is a constant,
which carries the correct dimension of the Lagrangian density in natural units.

%%%%%%%%%%%%%%%%%%%%%%%%%%%%%%%%%%%%%%%%%%%%%%%%%%%%%%%%
\subsection{One Real Scalar Field}

In the case of one real scalar field we have
\be
{\cal L}=\frac12\partial_{\alpha}\phi\partial^{\alpha}\phi-V(\phi)
\ee
The potential $V(\phi)$ speficies the system. As an example we consider
the model
\be
\label{pot12}
V(\phi)=\frac12\phi^2-|\phi|+\frac12
\ee
This potential was recently considered in Ref.~{\cite{the99}}. There are two
minima, at the values $\{v_1=1,v_2=-1\}$. The equation
of motion is $d^2\phi/dx^2=\phi-{\phi}/{|\phi|}$.
It has the solutions $\phi_{\pm}(x)=\pm({x}/{|x|})[1-\exp(-|x|)]$.
This model can be written in terms of the function
\be
W(\phi)=\phi-\frac12\,|\phi|\,\phi
\ee
This means that the topological sector is a BPS sector. The BPS solution
satisfies
\be
\frac{d\phi}{dx}=1-|\phi|
\ee
This solution is $\phi(x)=(x/|x|)\,[1-\exp(-|x|)]$. It has a kink-like
profile, and is linearly stable \cite{baz99}, minimizing the energy
to $E_B=1$.

%%%%%%%%%%%%%%%%%%%%%%%%%%%%%%%%%%%%%%%%%%%%%%
\subsection{Two Real Scalar Fields}

We exemplify the case of two real scalar fields by considering the
superpotential
\be
W(\phi,\chi)=\phi-\frac13\phi^3-r\phi\chi^2
\ee
where $r$ is a real parameter. Here the potencial gets to the form
\be
V(\phi,\chi)=\frac12(\phi^2-1)^2-r \chi^2+r(1+2r)\phi^2\chi^2+
\frac12 r^2\chi^4
\ee
There are two minima for $r<0$, $v_1=(1,0),v_2=(-1,0)$.
For $r>0$ thare are four minima, the two former ones, and also the other two
$v_3=(0,\sqrt{1/r}),v_4=(0,-\sqrt{1/r})$.

In this model the first order equations are
\ben
\frac{d\phi}{dx}&=&(1-\phi^2)-r\chi^2
\\
\frac{d\chi}{dx}&=&-2r\phi\chi
\een
We see that for $\chi\to0$ the first order equations demand
$d\phi/dx=(1-\phi^2)$, and the BPS defect solution is
$\phi(x)=\tanh(x)$.
However, for $\phi\to0$ the first order equations demand that $\chi^2=1/r$,
that is, that the $\chi$ field should be at the corresponding minima.
These results are manifestations that the sector defined by the minima
$(\pm 1,0)$ is a BPS sector, while the other sector, defined by the minima
$(0,\pm\sqrt{1/r})$ is a non-BPS sector, and the non-BPS solutions
are stable if and only if $1/r<1$ \cite{bbd00}.

The solution $\phi=\tanh(x)$ and $\chi=0$ is a one-field solution, but
the system also supports two-field solutions. In the sector that connect
the minima $(\pm 1,0)$, the two-field BPS solutions have the explicit form
\ben
\label{2sol1}
\phi(x)&=&\tanh(2r x)
\\
\label{2sol2}
\chi(x)&=&\pm \sqrt{\frac{1}{r}-2}\;{\rm sech}(2r x)
\een
which requires that $1/r>2$. The one-field and two-field solutions that
appear in the sector connecting the minima $(\pm 1,0)$ have the same energy,
$E=4/3$. In the $(\phi,\chi)$ space the real line ${x\in{\rm\bf R}}$ is
mapped to line segments: for the one-field solution we get a straight line
segment going from $(-1,0)$ to $(1,0)$, and for the two-field solutions
we get an ellyptic arc, obeying $\phi^2+\chi^2/(1/r-2)=1$.
These solutions map the Ising and Bloch walls that
appear in magnetic systems, respectively, and represent solutions of the
anisotropic $XY$ model, that describe interfaces between ferromagnetic
domains \cite{wal97,bbd00r}.

%%%%%%%%%%%%%%%%%%%%%%%%%%%%%%%%%%%%%%%%%%%%%%%%%%%%%%%%%%%%%%%%%%%%%%%%
\section{Extended Systems}
\label{sec:corrections}

Here we deal with systems defined by extended potentials, that differ from
the above primary potentials by some slight modifications. We start dealing
with the general model
\be
V_{\epsilon}=
\frac12\sum_{i=1}^nW^2_{\phi_i}+\frac12\;\epsilon\;F(\phi_1,...,\phi_n)
\ee
where $\epsilon$ is a real parameter, infinitesimal. In this case
we can use such parameter to control the procedure of extending the result
beyond the $\epsilon$-independent value. 
The correction to the potential depends on $F=F(\phi_1,...,\phi_n)$. In
general, we may have two different types of functions: functions that respect
and functions that do not respect the symmetry of the original system.
Both type of functions are important to describe situations where the system
is modified by the presence of external fields, chemical potentials, etc.
However, in the present work we are interested mainly on topological solitons,
in investigating the topological sectors of the model. Thus, we consider
the case of functions that respect some symmetry of the primary system.
In this case the function $F$ accounts for modifications of the original
system, without destroying the topological sector one is investigating.
This means that in the set of possible
vacuum states $\{v_a,v_b,v_c,...\}$ of the primary system, at least the
vacua $v_a$ and $v_b$ remain present, althought they can be slightly changed
to $v_a^{\epsilon}$ and $v_b^{\epsilon}$. We shall be
investigating slight modifications in the topological sectors of the BPS
type, modifications that do not destroy the sectors themselves. In this case
we write the static solution
$\phi^{\epsilon}_1(x),\phi^{\epsilon}_2(x),...,\phi^{\epsilon}_n(x)$
of the extended system in terms of the static solution
of the original model in the form, up to first order in $\epsilon$
\be
\label{phiepsilon}
\phi^{\epsilon}_i(x)=\phi_i(x)+\epsilon\,\eta_i(x),\;\;\;\;\;i=1,2,...,n
\ee

In the extended system, the energy of static solutions has the form
\be
E=\frac12\int^{\infty}_{-\infty}dx\sum_{i=1}^n
\Biggl[\left(\frac{d\phi_i}{dx}\right)^2+W^2_{\phi_i}\Biggr]+\frac12\,
\epsilon \int^{\infty}_{-\infty}dx\;F(\phi_1,\phi_2,...,\phi_n)
\ee
It can be written as, in the case of the BPS sector that connects the vacuum
states labeled by $a$ and $b$,
\be
E^{ab}=\frac12\int^{\infty}_{-\infty}dx\sum_{i=1}^n
\left(\frac{d\phi_i}{dx}-W_{\phi_i}\right)^2
+ E^{ab}_B+\frac12\,\epsilon \int^{\infty}_{-\infty}dx\;
F(\phi_1,\phi_2,...,\phi_n)
\ee
We now follow Ref.~{\cite{baz91}}. We see that the field
$\phi_i^{\epsilon}$ given by Eq.~(\ref{phiepsilon}) shows that the first
term in the above expression for the energy do not contribute to first
order in $\epsilon$. This fact allows writing the energy in the form
\be
E^{ab}=E^{ab}_B+\frac12\;\epsilon\int^{\infty}_{-\infty}dx\;
F(\phi_1,\phi_2,...,\phi_n)+{\cal O}(\epsilon^2)
\ee
Here $E^{ab}_B=|W_{ab}|$ and $W_{ab}=W(v_a^{\epsilon})-W(v_b^{\epsilon})$.
The correction to the potential is small,
and we separate two cases: first, the case where the correction does not
change the minima of the primary potential, that is, the case where
$v_a^{\epsilon}=v_a$ and $v_b^{\epsilon}=v_b$; and second, the case where the
correction slightly modifies the minima of the primary potential. In the
first case both $\phi_i^{\epsilon}(x)$ and $\phi_i(x)$ have the same
asymptotic behavior, thus $\eta_i(x)$ must vanish asymptotically;
in the second case $\phi_i^{\epsilon}(x)$ is asymptotically different
from $\phi_i(x)$, thus $\eta_i(x)$ cannot vanish asymptotically. In both
cases the term $E^{ab}_B$ exactly reproduces the corresponding term in the
primary system, up to first order in $\epsilon$.
This is so because
$W[\phi^{\epsilon}_1(\pm\infty),...,\phi^{\epsilon}_n(\pm\infty)]$
can be expanded to give
$$
W[\phi_1(\pm\infty),...,\phi_n(\pm\infty)]+\epsilon
\sum_{i=1}^n
\eta_i\;(\pm\infty)\;
\frac{dW}{d\phi_i}{\Bigg|}_{\phi_i(\pm\infty)}
$$
However, we know that $\phi_i(\pm\infty)$ are minima of the primary model,
and are extrema of W. Thus, the second term in the above expression vanishes.
For the topological charge, in the first case we see that it does not change,
giving $Q_T^{\epsilon}=Q_T$. However, in the second case the topological
charges change according to $Q_T^{\epsilon}=Q_T+\epsilon\;\Delta Q$, where
$Q_T^{\epsilon}$, $Q_T$ and $\Delta Q$ are $n$-component vectors, and
$\Delta Q$ accounts for the difference
$\eta_i(\infty)-\eta_i(-\infty), i=1,2,...,n$.  

The above results are general results, and we illustrate the general
procedure with some examples, spliting the investigation in the two
subsections that follows, which deal with one and
two real scalar fields separately.

%%%%%%%%%%%%%%%%%%%%%%%%%%%%%%%%%%%%%%%%%%%%%%%%%%%%%%%%
\subsection{The case of one real scalar field}

In the case of one field, let us first consider the $\phi^4$ model, defined
by the potential
\be
V(\phi)=\frac12(\phi^2-1)^2
\ee
We consider $F_1(\phi)=(\phi^2-1)^2$, which is
an example where the correction does not change the minima of the
primary system, and so $Q_T^{\epsilon}=Q_T=1$. In this case the energy
becomes
\be
\label{r11}
E=E_B\;\left(1+\frac12\;\epsilon\right)
\ee
where $E_B=(4/3)$. It can be greater $(\epsilon>0)$ or lesser
$(\epsilon<0)$ than the energy of the unperturbed system. 

We notice that the above $F_1(\phi)$ allows rewriting the potential as
\be
V(\phi)=\frac12(1+\epsilon)(\phi^2-1)^2
\ee
This potential requires that $\epsilon>-1$, and shows that the extended
system is very much like the primary model, with the coupling for
self-interaction changed by the $\epsilon$ term. In this case we can
find the energy of the static solution exactly. It is
\be
E=E_B\sqrt{1+\epsilon\;}
\ee
For $\epsilon$ very small, we expand the above result to get the former
answer, Eq.~(\ref{r11}), and this shows that our approach works correctly.

As a second example, let us consider the same primary model and another
function
\be
\label{12}
F_2(\phi)=(1-\phi^2)
\ee
In this case the correction does change the minima of the primary potential.
The new minima are at $\pm(1+\epsilon/4)$. The energy is
\be
\label{r12}
E=E_B\left(1+\frac34\,\epsilon\right)
\ee
It can be greater $(\epsilon>0)$ or lesser $(\epsilon<0)$ than the energy
of the unperturbed system. The topological charge changes to
$Q_T^{\epsilon}=Q_T\,(1+\epsilon/4)$.

Here we also notice that with the above correction of Eq.~(\ref{12}) the
potential can be rewritten in the form
\be
V(\phi)=\frac12\left(\phi^2-1-\frac12\epsilon\right)^2
\ee
which is correct to first order in $\epsilon$.
In this case the energy of the static solution is
\be
E=E_B\left(1+\frac12\epsilon\right)^{3/2}
\ee
However, since $\epsilon$ very small, we expand the above result to get the
former answer, Eq.~(\ref{r12}). We notice that the modification
in the minima of the potential changes the topological charge, and also
the energy of the topological solution. The energy increases or decreases,
depending on the increasing or decreasing of the spontaneous symmetry
breaking parameter.

We now consider the model defined in Eq.~(\ref{pot12}). We extend this model
with the above $F_2(\phi)$. The new potential is
\be
V_{\epsilon}(\phi)=\frac12\phi^2-|\phi|+\frac12+\frac12\epsilon(1-\phi^2)
\ee
For $\epsilon\neq0$, small, the minima change from $v_{\pm}=\pm1$ to
$v^{\epsilon}_{\pm}=\pm(1+\epsilon)$. The energy of the topological solution
changes to
\be
\label{newe}
E=1+\frac32\epsilon
\ee
We see that the energy decreases when the spontaneous symmetry breaking
parameter decreases.

We notice that the above potential can be written as
\be
V(\phi)=\frac12\left(\sqrt{1+\epsilon}-
\frac{|\phi|}{\sqrt{1+\epsilon}}\right)^2
\ee
This result is valid up to first order in $\epsilon$. It allows introducing
a superpotential, and the first order equation is
\be
\frac{d\phi}{dx}=\sqrt{1+\epsilon}-\frac{|\phi|}{\sqrt{1+\epsilon}}
\ee
It has the BPS solution
\be
\phi(x)=(1+\epsilon)\frac{x}{|x|}\left(1-e^{-|x|/(1+\epsilon)}\right)
\ee
This shows that both the amplitude and width of the former kink-like solution
change in the extended model. The energy is $(1+\epsilon)^{3/2}$, but
since $\epsilon$ is small we can expand this result to get the
former answer, given by Eq.~(\ref{newe}).

%%%%%%%%%%%%%%%%%%%%%%%%%%%%%%%%%%%%%%%%%%%%%%%%%%%%%%%%%%%%%
\subsection{The case of two real scalar fields}

The case of two fields is more involved, and we envisage several distinct
possibilities of illustrating this situation. We consider the example
presented in Sec.~{\ref{sec:general}}. We extend that model with the function
\be
\label{f1}
F_1(\phi,\chi)=(1-\phi^2)
\ee
In this case, in the BPS sector defined by the minima $(\pm1,0)$
the correction to the one-field solution is
\be
E^1_1=E_B\left(1+\frac34\epsilon\right)
\ee
In the case of the two-field solution $(\ref{2sol1})$ and $(\ref{2sol2})$
we get
\be
E^1_2=E_B\left(1+\frac38\,\epsilon\,\frac{1}{r}\right)
\ee
We introduce the ratio between energies, $R=E_2/E_1$. We see that
\be
R_1=1+\frac38\,\epsilon\,\left(\frac{1}{r}-2\right)
\ee
We consider another perturbation
\be
F_2(\phi,\chi)=r \chi^2
\ee
It gives
\be
E^2_1=E_B
\ee
and
\be
\label{f2}
E^2_2=E_B\Biggl[1+\frac38\,\epsilon\,\left(\frac{1}{r}-2\right)\Biggr]
\ee
They give
\be
R_2=1+\frac38\,\epsilon\,\left(\frac{1}{r}-2\right)
\ee
We see that $R_1=R_2$. The ratio $R=E_2/E_1$ does not depend on the way
one extends the model, using $F_1=(1-\phi^2)$ or $F_2=r\chi^2$. We notice that
the extension with $F_1=(1-\phi^2)$ changes the minima of the primary model
from $(\pm1,0)$ to $(\pm1\pm\epsilon/4),0)$. Thus, the topological charge of
the sector changes according to
\be
Q_T={1\choose0}\;\to\;Q^{\epsilon}_T=Q_T+\epsilon {{1/4}\choose{0}}
\ee
The other extension, that uses $F_2=r\chi^2$,
does not change the minima $(\pm1,0)$, so the topological charge in this BPS
sector remains the same $Q^{\epsilon}_T=Q_T$.
The above examples show two distinct ways of removing the degeneracy between
the standard one-field solution and two-field solutions (\ref{2sol1})
and (\ref{2sol2}). The two-field solutions are less energetic for
$\epsilon<0$. This means that one favors the non-trivial two-field
configuration when the symmetry breaking parameter decreases in one
or in the two $\phi$ and $\chi$ directions.

There is another way of removing the degeneracy of the one-field and the
two-field solutions that appear in the sector connecting the minima
$(\pm 1,0)$. We consider the case where the extra piece contains interactions
between the two fields, for instance
\be
F^{(k)}_3(\phi,\chi)=r\phi^{2k}\chi^2,\;\;\;\;k=1,2,...
\ee
In this case the minima of the primary system do not change, so the
topological charge gets no modification. The same happens to the energy
of the one-field solution. However, the energy of
the two-field solutions changes to
\be
E^{(k)}=E_B+\frac12\epsilon\left(\frac{1}{r}-
2\right)\frac{1}{2k+1}
\ee
We see that in the limit $k\to0$ ones obtains the former result,
giving in Eq.~(\ref{f2}). We also see that the sign of $\epsilon$ makes the
energy density of the two-field solutions to be higher $(\epsilon>0)$ or
lower $(\epsilon<0)$ than the energy density of the one-field solution,
removing the degeneracy they have in the primary system. 

In the former model of two real scalar fields with the function
$F_1(\phi,\chi)$ as in Eq.~(\ref{f1}), the potential has the form
\ben
V_{\epsilon}(\phi,\chi)&=&\frac12(\phi^2-1)^2-r\chi^2+
\frac12 r^2\chi^4\nonumber\\
& &+r(1+2r)\phi^2\chi^2+\frac12\epsilon(1-\phi^2)
\een
There are four minima, two at $(0,\chi_{\pm}),\;\chi_{\pm}=\pm\sqrt{1/r}$,
and two at $(\phi_{\pm},0),\;\phi_{\pm}=\pm(1+\epsilon/4)$. We compare this
potential with the potential of the primary model. The presence of the extra
term shows that, for $r>1$ the stable non-BPS solution implies
\be
V_{\epsilon}(\phi,0)=\frac12(\phi^2-1)^2+\frac12\epsilon(1-\phi^2)
\ee
and
\be
V_{\epsilon}(\phi,\pm\sqrt{1/r})=\frac12\phi^4+
\left(2r-\frac12\epsilon\right)\phi^2+\frac12\epsilon
\ee
This shows that the (squared) mass of the $\phi-$meson is, inside the
$\chi-$kink
\be
m^2_{\phi}(in)=4\left(1+\frac12\epsilon\right)
\ee
Outside the $\chi-$kink we get $m_{\phi}^2(out)=4r-\epsilon$, which
gives the ratio
\be
\frac{m^2_{\phi}(in)}{m^2_{\phi}(out)}=
\frac1r\Biggl[1+\frac12\left(1+\frac1{2r}\right)\epsilon\Biggr]
\ee
Thus, if the non-BPS $\chi-$kink entraps $\phi-$mesons in the primary system,
the entrapment is still more efficient in the extended system, for
$\epsilon<0$. We see that deviations from the BPS bound may improve
the efficiency of the mechanism for the entrapment of the other field. 

%%%%%%%%%%%%%%%%%%%%%%%%%%%%%%%%%%%%%%%%%%%%%%%%%%%
\subsection{Another case}

Let us now consider the model investigated in Ref.~{\cite{bbd00r}}. It is
described by three real scalar fields, and the potential has the form
\be
\label{pot3}
V(\sigma,\phi,\chi)=\frac23\left(\sigma^2-\frac94\right)^2+
\left(r\sigma^2-\frac94\right)(\phi^2+\chi^2)
+(\phi^2+\chi^2)^2-\phi(\phi^2-3\chi^2).
\ee
The projection with $(\phi,\chi)\to(0,0)$ gives
\be
V(\sigma)=\frac23\left(\sigma^2-\frac94\right)^2
\ee
This potential can be written with the superpotential
$W(\sigma)=(3\sqrt{3}/2)\sigma-(2\sqrt{3}/9)\sigma^3$. This fact shows
that the defect $\sigma(x)=(3/2)\tanh(\sqrt{3}x)$ that appears in this
case is a BPS defect. We can then extend this model adding to the
potential in Eq.~(\ref{pot3}) a term depending on the $\sigma$ field,
for instance
\be
f(\sigma)=\frac12\epsilon\left(\frac32-\sigma^2\right)
\ee
We use former results to see that this term contributes to decreasing
or increasing the energy of the basic defect, increasing or decreasing the
efficiency of the mechanism for the entrapment of the network the other
two fields $\phi$ and $\chi$ may generate.

%%%%%%%%%%%%%%%%%%%%%%%%%%%%%%%%%%%%%%%%%%%%%%%%%%%%%%%%%%%%%%%%%%%%%%
\section{Comments}
\label{sec:comm}

In this paper we have examined systems described by real scalar fields, in
which the energy of static field configurations is in the vicinity of the
BPS bound. This bound is attained by field configurations
that solve first order equations, and minimize the energy.
The BPS bound appears in the real bosonic sector
of supersymmetric theories described by chiral superfields. The systems we
have investigated are extensions of primary systems, described by potentials
given by the functions $W=W(\phi_1,\phi_2,...)$ and $F=F(\phi_1,\phi_2,...)$,
in the specific form
$$
\frac12W^2_{\phi_1}+\frac12W^2_{\phi_2}+
\cdots+\frac12W^2_{\phi_n}+\frac12\;\epsilon\;F(\phi_1,\phi_2,...,\phi_n)
$$
The parameter $\epsilon$ is real, and is used to control the deviation
from the primary system, described in terms of the superpotential $W$.

In the extended system, if the static solutions are
similar to the solutions one finds in the primary system, that is, for
$\phi_{i}^{\epsilon}(x)=\phi_{i}(x)+\epsilon\,\eta_i(x)$, we could examine
the energy related to the new defect solutions $\phi^{\epsilon}_{i}(x)$,
and write the first corrections in $\epsilon$ in a closed form, independently
of the specific form of the new defect solution itself. The formal results
are of direct interest to field theory, where they may
be used to improve the mechanism for the entrapment of
the other field \cite{b74,r75,m76,w85,m95,m95a}.

We have investigated specific systems, and we have found diverse
possibilities of removing the degeneracy between different types of
solutions, without destroying the degeneracy of the vacuum states. The
examples we have presented serve to illustrate some practical possibilities
of removing defect degeneration, and this is of direct interest in
application in specific physical situations. In ferroelectric crystals, for
instance, the order parameters that control structural phase transitions
may be changed by applications of external pressure along specific planar
directions in the crystal. This is a tipical scenario for changing the
parameters that control spontaneous symmetry breaking inside the crystalline
material, changing the energetics of the structural phase transition. In the
systems we have examined in the present work, the presence of external
pressure may be directly mapped into specific forms of $F$, that controls
the extended system. The present work opens a new route for exploring systems
of coupled scalar fields, intending to mimic specific systems
in applications to Cosmology and to Condensed Matter. We postpone to the
near future some specific investigations.

\acknowledgements
We would like to thank J R S Nascimento and R F Ribeiro
for discussions. C A G Almeida thanks Departamento de Matem\'atica,
Universidade Regional do Cariri, 63.100-000 Crato, Cear\'a, Brazil,
for support. This work is supported in part by funds provided by CAPES,
CNPq, and PRONEX.

\end{document}